Practical Considerations for Finite Concentrations Molecular Dynamics Simulations

Xiaoxu Ruan[1], Fabrice Roncoroni[2], David Prendergast[2] and Tod A Pascal[1*]

1. ATLAS Materials Physics Lab, Aiiso Yufeng Li Family Department of Chemical and Nano Engineering, University of California, San Diego, La Jolla, CA 92093
2. Theory of Nanostructured Materials Facility, Molecular Foundry, Lawrence Berkeley National Lab, Berkeley, CA 94720

## Abstract

Understanding concentrated electrolytes requires a theory that spans local hydration and mesoscale interfacial assembly. We present an integrated workflow–SCOPE–that combines (i) enhanced sampling focused on a single $Li^+$ ion, (ii) reweighting of biased trajectories to recover equilibrium microstate probabilities, and (iii) a chemical-potential correction that accounts for the limited reservoir of free water in finite simulation boxes. Applied to $LiCl_{(aq)}$ across 0.5–26 M and 283–313 K, this approach reveals a simple organizing principle: solvated ions dominate at low concentration; contact ion pairs emerge at intermediate strength; and aggregated Li-xCl clusters become most stable at the solubility limit. The resulting free-energy trends predict temperature-dependent solubility in close agreement with experiment and clarify the role of interfacial nucleation in precipitation. Beyond the simple $LiCl_{(aq)}$ salt considered here, SCOPE offers a transferable strategy for characterizing speciation and phase behavior in concentrated liquid systems where collective coordinates and rare events dominate.

## Introduction

The microstructure of a liquid system governs its macroscopic properties, including solubility, phase stability, and transport behavior. Solutes and ions in water organize over disparate length scales. At short distances, various intermolecular forces, such as Pauli exclusion/repulsive packing, electrostatic, van der Waals, and hydrogen bonding, determine local structure, while at longer distances, interfaces and clusters mediate assembly.[1–4] This dichotomy becomes central in concentrated electrolytes, where solvent reorganization is frustrated by ion crowding and collective motifs govern thermodynamics and transport. Despite extensive experimental studies on solubility, precipitation, and phase stability, the molecular-level understanding of these processes remains largely incomplete.[5] This is because while experimental techniques provide direct measurements of macroscopic thermodynamic properties, they typically lack molecular-level resolution, making it difficult to directly link solvation structures to bulk behavior.

Computational approaches, such as molecular dynamics (MD) simulations, offer an atomic-scale perspective and direct observation of solvation structures and their fluctuations. However, several challenges remain in simulating finite concentrations beyond the infinite dilution limit in these simulations. First, traditional metrics, like radial distribution functions and coordination numbers, average over microstate diversity and can therefore obscure the states that control phase behavior[6] – especially near precipitation, where rare structural transitions drive the onset of insolubility. A more incisive view is to analyze the ensemble of trajectories that produce these transitions, favoring operational quantities (microstate probabilities, free energies) over averages that blur heterogeneity. While previous works have been shown to somewhat mitigate artificial clustering and concentration-dependent biases, they often rely on a uniform correction factor that does not explicitly consider molecular-level variations in free solvent availability, which, we hypothesize, is critical for accurate free energy calculations.[7–10]

Second, MD is only as good as its sampling. If trajectories fail to explore the relevant regions of phase space, any thermodynamic quantity derived from them is suspect. This risk grows acute at high concentration, where rugged free-energy landscapes trap the system in local minima. Under these conditions, the outcome depends on the starting geometry–a violation of ergodicity that renders ensemble averages meaningless. Classical MD, constrained by nanosecond timescales, rarely captures the rare events that matter most: the structural rearrangements that nucleate precipitation and phase segregation near the solubility limit.

To overcome this barrier, one must accelerate the exploration of configuration space without guessing the reaction coordinate. Metadynamics[11,12] does exactly that: it adds a history-dependent bias along chosen collective variables, gradually filling free-energy wells and pushing the system toward new regions. By doing so, metadynamics converts inaccessible transitions into observable ones, revealing molecular precursors to phase change—details that conventional MD leaves hidden behind kinetic bottlenecks.

This paper adopts an expanded perspective for considering the local microstructural changes in aqueous LiCl solutions, as the ionic strength increases. We integrate enhanced sampling, trajectory-space reweighting, and a chemical-potential correction grounded in water activity to recover an equilibrium free-energy spectrum over distinct solvation microstates, from dilute solutions to supersaturation. The physical picture that emerges mirrors canonical ideas about water and assembly: at small scales, hydration constraints impart entropic costs; at mesoscopic scales, interfaces nucleate and stabilize aggregates; the crossover in stability is concentration- and temperature-dependent. We show that this organizing principle rationalizes speciation across 283–313 K and predicts solubility limits consistent with experiment.

## Methodology

We developed a computational workflow (**Fig. 1**) for elucidating the unique solvation states in MD simulations as a function of finite concentration: SCOPE – **S**olvation **C**haracterization via **O**ptimized **P**robability **E**nsemble averaging – comprising the following features

1. <u>Classical MD Simulation Setup & System Details</u>

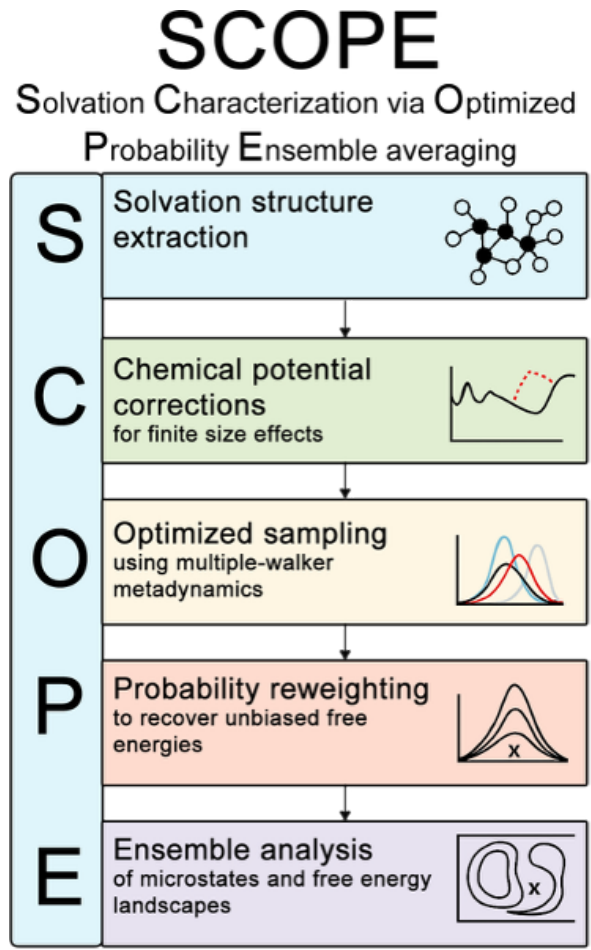

**Figure 1. SCOPE workflow.** From classical MD to single-ion metadynamics and trajectory-space reweighting, with a finite-reservoir chemical-potential correction. The organizing principle—hydration → pairing → aggregation—emerges from corrected free-energy spectra across concentration and temperature.

Atomistic MD simulations of aqueous LiCl solutions at varying concentrations were simulated using the LAMMPS MD code.[13] We conducted simulations on different molarities of LiCl solutions, from dilute to beyond the solubility limit at three different temperatures: from 0.5 M – 23 M at 283 K, from 0.5 M – 25 M at 298 K, and from 0.5 M – 26 M at 313 K. All simulations employed the SPC/E water model,[14] while the ions were described by the AMBER force field,[15] which was specifically optimized using hydration free energies of solvated ions, lattice energies, and lattice constants of alkali halide salt crystals. Simulations of various numbers of $Li^+$ and $Cl^-$ ions, together with the appropriate number of water molecules (**Table S1**) were performed under periodic boundary conditions. The van der Waals interactions between atoms were modeled using a Lennard-Jones potential with a real-space cutoff of 10 Å. Short-range electrostatic interactions used the same cutoff, while long-range electrostatics beyond the cutoff were treated using the particle-particle particle-mesh (PPPM) method,[16] with a convergence tolerance of $10^{-5}$.

In each system, the initial atomic configuration was first relaxed using the conjugate gradient energy-minimization algorithm with a force and energy convergence tolerance of $10^{-4}$ kcal/(mol·Å) and $10^{-4}$ kcal/mol, respectively. Subsequently, we annealed the system by heating to 400 K for 10 ps, then back down to the target temperature, within the canonical (i.e., constant number of particles, volume, and temperature, or NVT) ensemble. The temperature was controlled by a Nosé-Hoover thermostat with a relaxation time of 100 fs. The system was then equilibrated under an isothermal-isobaric (NPT) ensemble at a pressure of 1 atm for 1 ns. We controlled the pressure by means of a Nosé-Hoover barostat, with a relaxation time of 2 ps. The equations of motion used are those of Shinoda et al.[17] with a numerical timestep of 1 fs, which combines the hydrostatic equations of Martyna et al.[18] with the strain energy proposed by Parrinello and Rahman[19]. The time integration schemes closely follow the time-reversible measure-preserving Verlet integrators derived by Tuckerman et al.[20]. The equilibrium cell parameters were averaged over the final 250 ps. To ensure a well-defined simulation box, the system was then deformed to match its equilibrium size over 100 ps, followed by an additional 500 ps NVT equilibration to remove any residual stress before production runs. The final production simulations were conducted under NVT ensemble at the target temperatures for 5 ns. The system's thermodynamic and structural properties were monitored throughout the simulation.

2. <u>Advanced Sampling via Metadynamics</u>

To enhance sampling of solvation structures, we employed metadynamics (MetaD) using the COLVARS module [21] implemented in LAMMPS. The initial configurations (atomic positions and velocities) were the final snapshot of the equilibrium classical MD simulations above. Three collective variables (CVs) were used to bias the system: the Li-O, Li-Cl, and Li-Li coordination numbers, where the coordination numbers were between a single $Li^+$ and the different groups of atoms, described by **Eqn 1**: [22]

$$CN(Li^+, group1) = \sum_{i \epsilon group1} \frac{1-(|x_{Li}-x_i|/d_0)^n}{1-(|x_{Li}-x_i|/d_0)^m} \quad (1)$$

where group1 is the group of atoms, cutoff $d_0$ is the "interaction" distance (Å), $n$ is the numerator exponent for the switching function, and $m$ is the denominator exponent for the switching function. Each coordination number had an upper boundary of 7.0 and a lower boundary of 0.0. The cutoff distances $d_0$ were chosen as Li-O: 2.65 Å, Li-Cl: 2.95 Å, and Li-Li: 5 Å, determined based on the first solvation shell radius from the classical MD radial distribution function (RDF).

Notably, the Li-Li cutoff distance was set to a larger value to account for Li-Li clustering effects. To better isolate the biased potential for each sampled cluster while still considering the concentration effect, we applied the biased potential to only one $Li^+$ in each system. This approach simplifies the free energy calculation in the subsequent data analysis.

The free energy surface was stored on a discrete grid with a spacing of 0.1. A biasing potential was applied every 1000 simulation steps by adding a Gaussian hill potential, with a hill weight of 0.6 kcal/mol and a hill width of 1.25. To further enhance sampling of solvation structures, we employ multiple-walker metadynamics,[23] with 10 walkers communicating every 25 ps, which enhances configurational space exploration by applying a history-dependent bias potential. To accelerate the initial exploration of the free energy landscape, different restart configurations were selected from equilibrated classical MD simulations, allowing each walker to begin from a distinct region of phase space and ensuring more efficient sampling.

### 3. Solvation Structure Extraction and Cluster Identification

To extract microscopic insights beyond ensemble averages, we adopt a workflow inspired by the algorithm developed by Roncoroni et al.,[6] in which solvation structures are extracted from MD trajectories and analyzed via dimensionality reduction, unsupervised clustering, and clustering alignment. In this work, clusters are constructed using only $Li^+$-centered distance cutoffs of Li–O = 2.65 Å and Li–Cl = 3.05 Å, obtained also from RDF (with a small buffer for connectivity). Since we didn't include Cl-H distance cutoff, our primary extraction rules are therefore $Li^+$-centered and do not explicitly encode water-mediated connectivity to $Cl^-$. As a result, configurations that would conventionally be termed solvent-separated ion pairs (SSIP) are not resolved as a distinct microstate at the extraction stage and are instead grouped into our coarser "solvated ions (SI)" category in the main analysis. This approach enables the identification of distinct geometries, even for species that share identical chemical formulas, thereby offering a more nuanced view of solvation behavior. Because of the relative simplicity of the LiCl(aq) system and limited variety of its coordination environments, only the first part of the algorithm was used to extract solvation structures from the MD trajectory. However, the analysis presented in this manuscript can readily be expanded to include the full workflow. This aspect will be essential for analyzing more complex systems where conventional metrics fail to resolve structural heterogeneity.

### 4. Free Energy Calculations

We define each unique solvation structure as a microstate:

$$Z = \Sigma_i e^{-\beta E_i} \ (\beta = \frac{1}{k_B T}) \quad (2)$$

where $Z$ is the partition function, which serves as a normalization constant and encapsulates the statistical weight of all possible states, $i$ is the index over all discrete microstates, $E_i$ is the relative free energy of the microstate, $\beta$ is the inverse thermal energy, $k_B$ is Boltzmann's constant, and $T$ is the absolute temperature in Kelvin. From $Z$, we can write:

$$P_i = \frac{e^{-\beta E_i}}{Z} \quad (3)$$

where $P_i$ is the probability of finding the system in state $i$ at equilibrium. This equation expresses a Boltzmann distribution, indicating that lower-energy states are more probable. From this, we can define:

$$A_i = -k_B T \ln P_i \quad (4)$$

where $A_i$ is the Helmholtz free energy. This equation allows us to compute the relative energy associated with each family of structures directly from their observed or calculated probabilities.

In enhanced sampling techniques such as metadynamics, a time-dependent bias potential $V(s,t)$ is applied along selected collective variables $s$ to accelerate sampling. As a result, the observed probabilities $P_{unbiased}(i)$ do not reflect the true equilibrium distribution. To recover the unbiased probability $P_{unbiased}(i)$, a reweighting procedure can be employed:[24,25]

$$P_{unbiased}(i) = \frac{P_{biased}(i) e^{\beta V(s_i,t)}}{\Sigma_j P_{biased}(j) e^{\beta V(s_j,t)}}, \quad (5)$$

where $P_{unbiased}(i)$ is the probability of microstate $i$ observed in the biased simulation, $V(s_i,t)$ is the total bias potential applied at the collective variable value $s_i$ at time $t$ and the denominator ensures normalization over all sampled microstates $j$. This equation corrects the distorted probabilities caused by the bias potential, allowing accurate estimation of thermodynamic properties such as free energies. Once the unbiased probabilities $P_{unbiased}(i)$ are obtained, they can be used in Eqn. 4 to reconstruct the free energy (Helmholtz free energy) landscape. The stability of solvation structures is then analyzed as a function of concentration and corrected free energy values.

5. Correction for Finite-Size Effects

To account for the aforementioned effects due to the finite water reservoir beyond the infinite dilution limit, we introduce a chemical potential correction based on activity coefficients. Since the simulation box is much smaller than a real system, the reservoir of free water is limited, which can lead to an overestimation of the chemical potential when water molecules transition between the solvated and bulk phases. To account for this finite-size effect and ensure consistency with the actual chemical potential of water in the system, the computed cluster free energy must be adjusted accordingly.

We use the following equations to correct the chemical potential:

$$\mu_i = \mu_i^0 + k_B T \ln(\gamma x_i) \quad (6)$$

where $\gamma$ is the activity coefficient of free water in the solution, $\mu_i$ is the chemical potential of species $i$ in the system, $\mu_i^0$ is the standard chemical potential of species $i$, $x_i$ is the mole fraction of species $i$ in the solution. Notably, the activity coefficient $\gamma$ accounts for non-ideal effects due to temperatures and finite concentrations. In an ideal solution, $\gamma$ is equal to 1, so $\gamma x_i$ reduces to the mole fraction $x_i$. In real solutions, interactions with the Li$^+$ and Cl$^-$ ions disrupt the water structure, changing its local chemical potential, and the activity coefficient $\gamma$ corrects for this deviation from ideality. Specifically, we have:

$$x_{free\ water} = \frac{N_{free\ water}}{N_{water}+N_{LiCl}} \quad (7)$$

where $x_{free\ water}$ is the mole fraction of "free" (unbound) water molecules in solution, $N_{free\ water}$ is the number of free water molecules, $N_{water}$ is the total number of water molecules in the system, and $N_{LiCl}$ is the number of salt molecules (or potential ion pairs) in the system. This equation accounts for how LiCl affects the water environment by binding some water molecules. From this, we write:

$$\Delta\mu = k_B T ln(\frac{\gamma x_{free\ water}}{x_{bulk}}) \quad (8)$$

and

$$\Delta A_{corrected} = \Delta A_{cluster} - N_{water\ in\ cluster} \times \Delta\mu \quad (9)$$

where $\Delta A_{corrected}$ is the corrected free energy after considering the water activity, $\Delta A_{cluster}$ is the free energy of the cluster obtained from the MD simulations, and $N_{water\ in\ cluster}$ is the number of water molecules present in the cluster. **Eqn. (8)** and **(9)** correct the cluster free energy ($\Delta A_{cluster}$) by subtracting the contribution to the chemical potential due to the bound water molecules. The saturated LiCl solution water activity coefficient at 298 K is around 0.113, and data at other ionic strengths were adapted from experimental data for LiBr solutions,[26] (**Table S2** and **Fig. S3**). A similar procedure was used to obtain $\gamma$ at 283 K and 313 K (**Fig. S4**).

Finally, since the bias potential was applied to only one Li$^+$ in the system, we refined the method of selecting free water molecules – i.e., those not bound to any other Li$^+$ or Cl$^-$ ion –as being within a 12 Å radius around the central Li$^+$ ion. This selection ensures that the subsequent analysis focuses on the local solvation environment directly influenced by the biased ion.

6. Bootstrap Analysis and Statistical Analysis

To increase the statistical reliability of the data, we performed a bootstrapping resampling technique, where the data from a single trajectory is sampled from different starting frames (frame 0-9) within a fixed sampling interval (i.e., 10 frames), from which we calculated the mean and variance of the corrected free energy values.

## Results and Discussion

Our corrected free-energy spectra reveal a simple but powerful organizing principle for concentrated LiCl$_{(aq)}$: at low concentration, solvated-ion (SI) dominate; at intermediate concentration, contact ion pairs (CIP) compete with SI; near and beyond the solubility limit, aggregated Li–xCl clusters (AGG) become the most stable species. This tripartite ordering is robust across temperatures (283–313 K) and insensitive to moderate variations in metadynamics parameters when the bias is applied to a single Li$^+$ ion.

1. <u>Solvation Structures in Dilute LiCl Solutions</u>

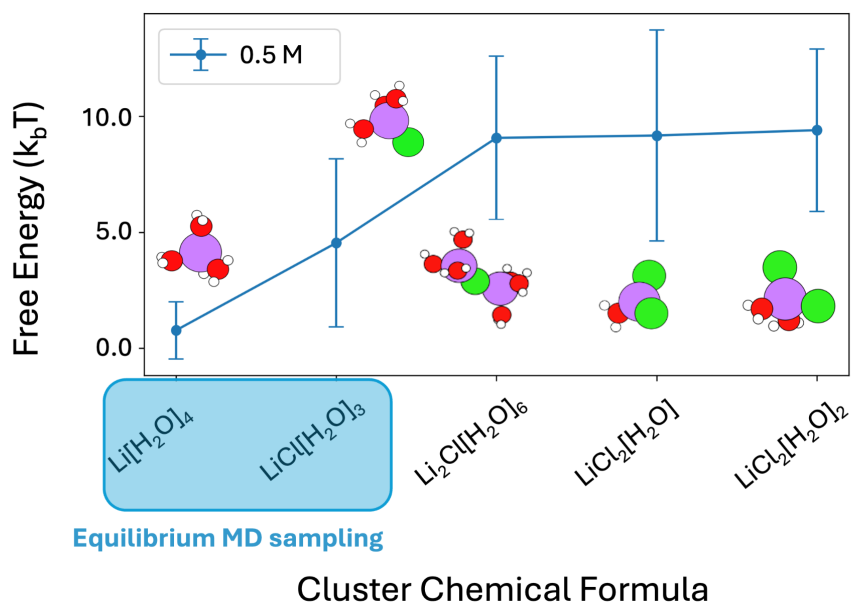

**Figure 2**. **Dilute LiCl$_{(aq)}$ at 298 K.** Top-five species from SCOPE. Fully solvated Li[H$_2$O]$_4$ (SI) dominates; LiCl[H$_2$O]$_3$ (CIP) is the principal minority. Bridge (BRG) and aggregate (AGG) motifs appear with low probability under enhanced sampling.

Metadynamics simulations provide much more efficient sampling, allowing the capture of dynamic structural changes and the identification of metastable solvation structures that may not appear in classical MD. For example, one can compare the equilibrium MD solvation structures in **Fig. S1** to the result in **Fig. 2**. More specifically, in dilute solutions (e.g., 0.5–1 M), the most probable microstate is the SI, typically represented as Li[H$_2$O]$_4$ with a full first shell of water molecules. The next-most probable state is the CIP LiCl[H$_2$O]$_3$, while bridging motifs (BRG) and small AGG rarely appear under equilibrium MD. Enhanced sampling exposes these rarities and places them appropriately high in the free-energy hierarchy. The increasing free-energy penalty for BRG and AGG at low ionic strength reflects the cost of disrupting hydrogen-bonded water without compensating interfacial stabilization.

2. <u>Solvation Structures Near and Beyond the Solubility Limit</u>

**Fig. 3** illustrates the solvation structures near and beyond the LiCl solubility limit at 298 K, and **Fig. 4** illustrates the free energy trend of solvation structures across different concentrations. Here we find as concentration rises into the mid-molar regime, SI loses stability relative to CIP and BRG (**Fig. 4**). The physical driver is entropic: pairing liberates waters from the first shell and reduces local constraints, leading to small decreases in A for CIP and BRG in intermediate regimes before both climb again at higher ionic strengths. In practice we observe a shallow minimum or plateau for CIP free energy as concentration increases, consistent with prior expectations that ion pairing becomes favorable when hydration shells crowd and overlap.

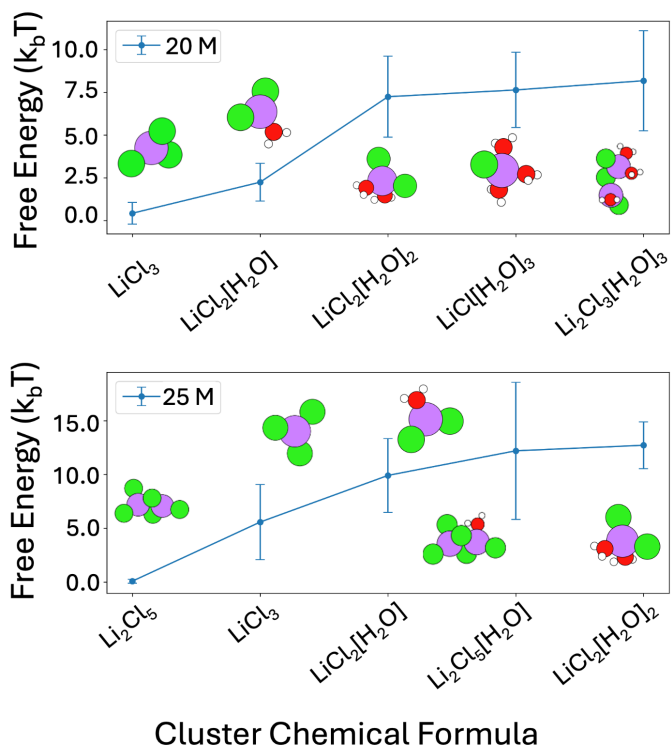

Cluster Chemical Formula

**Figure 3. Near and beyond solubility (298 K).** At ~20 M, Li–3Cl aggregates are most stable; at 25 M, larger conjugated clusters such as 2Li–5Cl dominate. Aggregation behaves as interfacial nucleation: mesoscale ionic interfaces stabilize clusters that drive precipitation.

Approaching the experimental solubility (~20 M[27]), the free energy of aggregated Li–xCl clusters drops sharply (**Fig. 4**). Microstates such as Li-3Cl become thermodynamically favored at the solubility limit, and larger, more conjugated aggregates like 2Li–5Cl dominate beyond it (e.g., 25 M) (**Fig. 3**). This crossover evidences interfacial nucleation: at high ionic strength, the collective coordination of Li and Cl builds mesoscale ionic interfaces whose surface area-to-volume tradeoffs favor compact clusters. The AGG free energy decreases from >17 $k_BT$ at 0.5 M to <2 $k_BT$ at ~20 M, crossing other microstates to become lowest at the solubility limit.

These results reinforce that the free energy surface reconstructed from metadynamics accurately captures relative stability trends without the need for computationally expensive quantum mechanical calculations. We propose that this efficiency and accuracy can be exploited as a general approach for predicting solubility limit and solution-phase speciation.

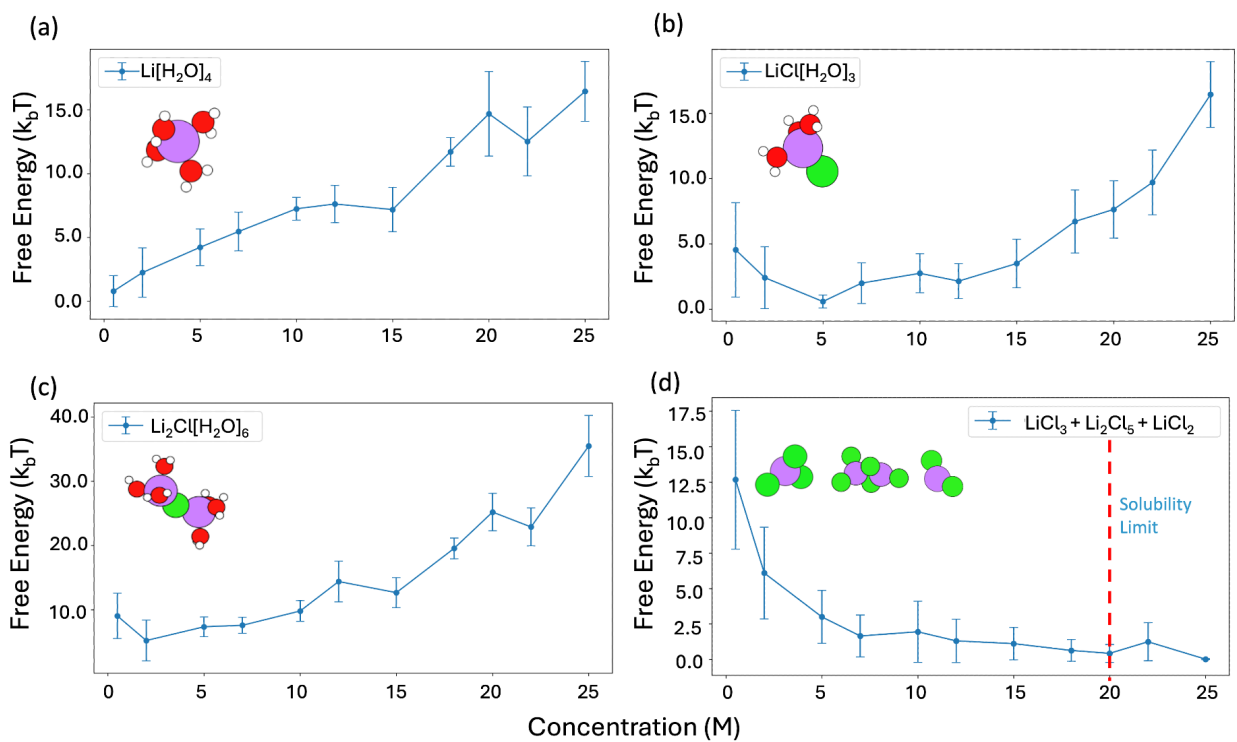

**Figure 4**. **Free energy vs. concentration (298 K).** SI, CIP, and BRG increase with ionic strength, while AGG drops sharply (from >17 $k_BT$ at 0.5 M to <2 $k_BT$ near 20 M), crossing to lowest free energy at the solubility limit.

Because CIP and BRG become briefly competitive with SI at intermediate concentrations, transport properties that depend on local coordination (e.g., ionic conductivity, viscosity) may exhibit non-monotonic behavior. Near solubility, the growth of AGG microstates suggests a decoupling between charge carrier number and mobility, consistent with prior observations that ion pairing and clustering reduce effective charge transport relative to Nernst–Einstein estimates.[28–30]

3.  <u>Predicting Temperature-Dependent Solubilities</u>

Across temperatures, the concentration-driven speciation picture persists, but temperature modulates the relative stability of SI and CIP (**Fig. 5**). At 313 K, the entropic penalty of constrained waters is higher, favoring direct Li–Cl pairing relative to full hydration; at 283 K, sluggish dynamics increase the uncertainty in AGG free energies at very high concentrations but do not change the crossover that places aggregates as the lowest free energy at the solubility limit (**Fig. 6**): ≈17 M at 283 K, ≈20 M at 298 K, and ≈21 M at 313 K. These trends are consistent with the view that hydration is entropy-limited at elevated temperatures while aggregate stabilization is dominated by collective coordination and incipient interfacial forces.

Fixed-charge force fields tuned for room-temperature thermodynamics can suffer from cancellation of energy and entropy errors at other temperatures. Our results—particularly the enhanced pairing at 313 K—could reflect genuine entropy-driven effects and/or limitations of the parameterization. For example, we have shown that room temperature optimization results in large cancellation of errors in the separate energy and entropy potentials, leading to incorrect temperature-dependent thermodynamic properties.[31] Future work will decompose the free energy into separate enthalpic and entropic terms (e.g., via temperature derivatives or two-state fits) to guide the reparameterization of polarization models.

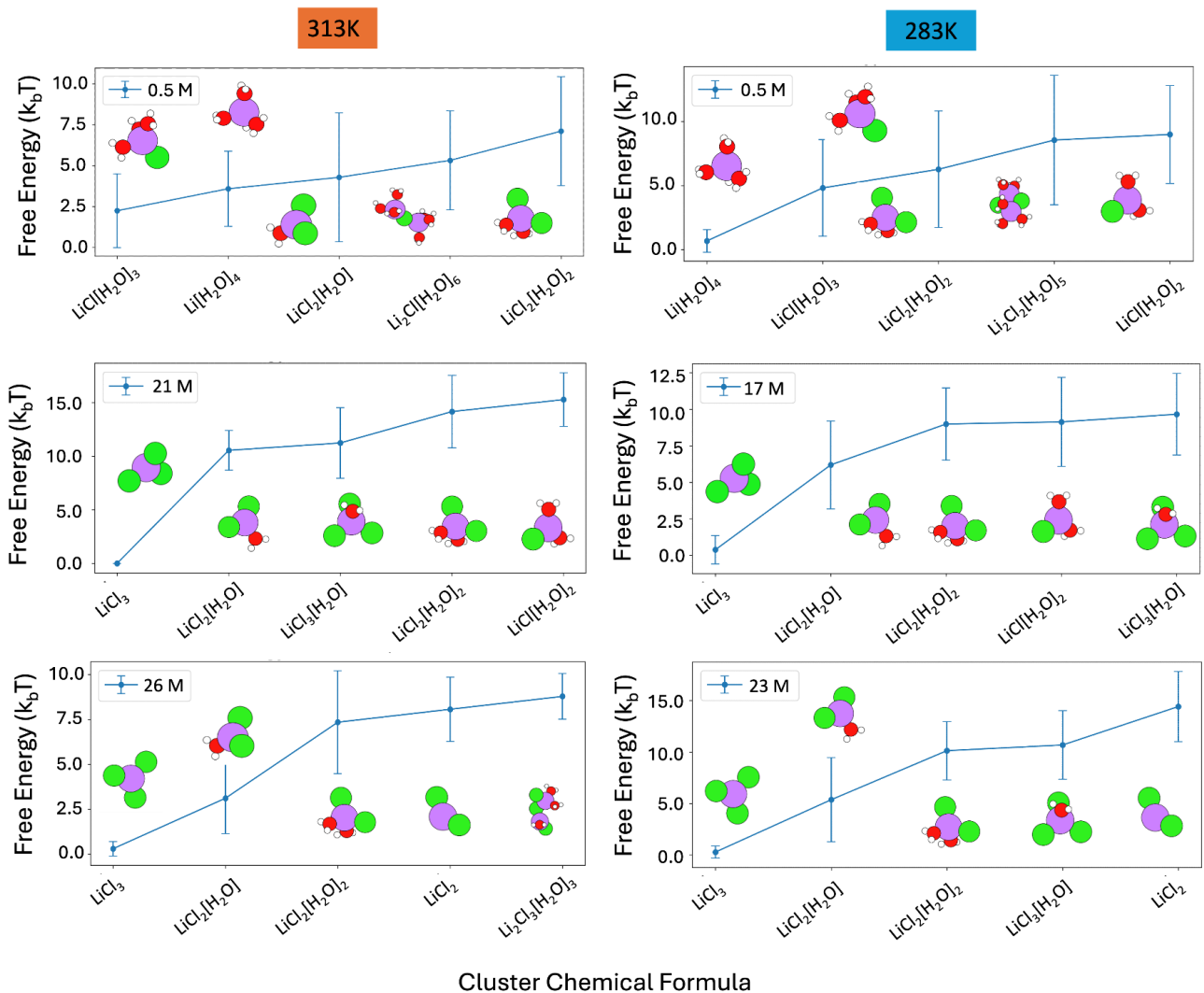

**Figure 5. Temperature dependence (283 and 313 K).** At low concentration, SI and CIP dominate; at each temperature's solubility limit, AGG becomes most stable. Elevated temperature favors ion pairing relative to full hydration due to the entropic penalty of constrained waters.

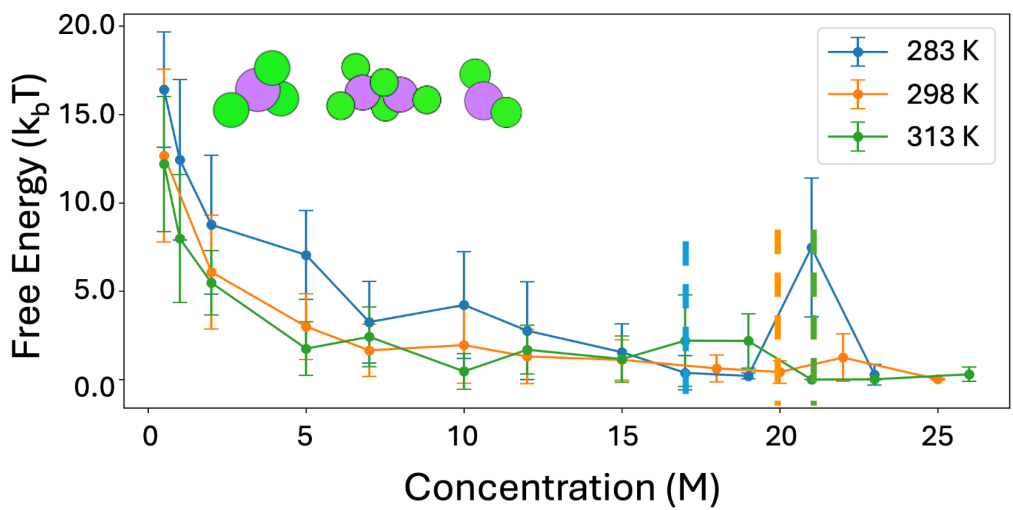

**Figure 6. Aggregate free energies across 283/298/313 K.** AGG is lowest at the solubility limit for each temperature (dashed lines). Low-temperature sluggish dynamics increase uncertainty for select points, but aggregate dominance persists.

4. The Effects of Chemical Potential Correction

The finite-reservoir correction—implemented via **Eqn. (8)** and subtracted for each bound water in the cluster (**Eqn. (9)**)—prevents artificial stabilization of oversized clusters (**Fig. S2**) caused by limited free water in small boxes. We defined $x_{free\,water}$ within a 12 Å sphere around the biased Li$^+$. Varying this radius by ±2 Å did not alter the crossover concentrations but changed the magnitude of the correction within uncertainty bounds. This local definition focuses the correction on the chemically relevant neighborhood of the biased ion and reduces sensitivity to distant, less-involved solvent molecules. Finite-size tests with varied box dimensions showed that the activity-based correction systematically damped size-dependent biases in cluster stability. Without this correction, the biased free-energy surface can overcount large AGG microstates; with the correction, the crossover to aggregate dominance appears cleanly at high concentration and aligns with experimental solubility limits. The method thus filters out finite-size artifacts by restoring bulk-relevant water activity around the biased ion.

5. A Length Scale Organizing Principle

While our analysis is based on microstate populations and free energies, the enhanced sampling trajectories provide qualitative insight into how the system moves between SI, CIP/BRG, and AGG families. In dilute regimes, transitions typically involve partial dehydration followed by direct $Li^+$–$Cl^-$ contact; in intermediate regimes, transient bridges form through shared anions; near solubility, multi-ion frameworks grow by sequential addition of $Cl^-$ to $Li^+$-centered motifs. Mapping these transitions onto pathway ensembles would enable kinetic rate predictions, and our single-ion bias approach is compatible with trajectory-space methods (e.g., transition path sampling)[32,33] that avoid presuming reaction coordinates.

## 6. Current Limitations and Future Extensions

A current limitation of the present framework is that activity coefficients are taken from experimental data rather than computed directly. While this enables quantitative free-energy estimates for $LiCl_{(aq)}$, it reduces the generalizability and predictive power of the method for conditions or chemistries where reliable measurements are unavailable. In future work, we will develop a fully computational route to estimate activity coefficients from molecular simulations—for example, by computing the osmotic pressure from MD using thermodynamic estimators and then deriving activity coefficients from the osmotic pressure.[34–36] Integrating such an approach would close the loop between simulation, solvation-state populations, and free-energy predictions, enabling end-to-end, data-independent solubility and thermodynamic modeling.

Another limitation is that our main analysis uses a coarse SI category (as defined in the Methodology), which groups together $Li^+$–$Cl^-$ configurations that are not in direct contact; as a result, SSIP is not explicitly represented as a distinct microstate. To assess whether this missing resolution matters, we performed a test at 0.5 M LiCl, using an alternative extraction in which we included both $Li^+$- and $Cl^-$-centered clusters and defined Cl-H distance cutoff 2.85 Å, and then subdivided SI into two classes: (i) solvent-separated ion pairs (SSIP), where $Li^+$ and $Cl^-$ are separated by exactly one water molecule, and (ii) isolated solvated ions (i-SI), where they are separated by two or more water molecules. This yields a higher-resolution solvation free-energy spectrum (**Fig. S5 (a)**). The resulting stability ordering is i-SI > SSIP / CIP > BRG, consistent with the 0.5 M results in **Fig. 2**.

Within a fixed composition such as $LiCl[H_2O]_9$, SSIP- and CIP-like geometries can coexist. Using a simple $Li^+$-$Cl^-$ distance criterion (3.2 Å cutoff), SSIP is dominant across most bootstrap samples and is predicted to be more stable than CIP within this composition (**Fig. S5 (b))**. This test case highlights that even simple systems like dilute $LiCl_{(aq)}$ can exhibit unexpectedly rich local structures diversity.[6] However, because this additional resolution does not improve prediction of the solubility limit, we retain the coarser chemical formula-based definition here. We expect that such finer partitioning, combined with a more well-defined, physically motivated classification scheme (e.g., SSIP, CIP, AGG), will be particularly valuable for chemically heterogeneous systems, such as localized high-concentration electrolytes, where this added granularity can aid interpretation of liquid-phase dynamics.[37]

Future work will consider more detailed trajectory analysis and advanced data-driven techniques, which may uncover more nuanced solvation modes and configurational transitions. Molecular liquids, with the significant thermal fluctuations and mobility of species in liquid systems, exhibit slow motions and long-timescale correlations that challenge both traditional sampling and quantum mechanical approaches. In these systems, the free energy landscape plays a dominant role in determining transport properties, underscoring the need for high-resolution solvation modeling. A more complete characterization of solvation and soft vibrational modes – spanning well-separated frequencies – will allow for the development of multidimensional free energy functionals, offering more in-depth insight into the dynamics of liquid systems. Such information is not only critical for understanding solvation thermodynamics but also valuable for developing machine learning models in drug discovery, where binding affinity depends not only on static energy but also on free energy contributions and dynamic fluctuations affecting ligand efficacy.[38]

## Conclusions

We have constructed an operational framework—SCOPE—for resolving speciation in concentrated electrolytes. The method combines enhanced sampling by metadynamics, trajectory-space reweighting, and a chemical-potential correction that restores bulk water activity in finite boxes. Applied to $LiCl_{(aq)}$, this approach captures the concentration-driven crossover in stability: hydration dominates at low ionic strength, ion pairing competes at intermediate regimes, and aggregated Li–Cl clusters become most stable near the solubility limit. These trends emerge without resorting to expensive quantum calculations, because the essential physics— length scales, interfaces, and rare events—is encoded in the free-energy spectrum reconstructed from biased trajectories.

By treating solvation structures as discrete microstates and building a partition function over their probabilities, we recover thermodynamic ordering and predict temperature-dependent solubilities in close agreement with experiment. The finite-reservoir correction filters out size artifacts that would otherwise spuriously stabilize large clusters, aligning simulation with chemical reality.

Beyond LiCl, the logic generalizes. Wherever concentrated liquids exhibit frustrated hydration and collective organization, SCOPE offers a scalable route to quantify speciation and phase behavior. Its ingredients—enhanced sampling, trajectory-space analysis, and activity-based corrections—are transferable to problems from ion transport and catalysis to drug formulation and soft-matter assembly, where solvation governs kinetics and stability. In short, by focusing on the dominant physics and sampling the rare events that matter, we turn molecular dynamics into a predictive tool for complex liquid systems.

## Acknowledgements

This work was primarily supported by the Energy Storage Research Alliance "ESRA" (DEAC02-06CH11357), an Energy Innovation Hub funded by the US Department of Energy, Office of Science, Basic Energy Sciences. Theoretical portions of work were performed as a user project at the Molecular Foundry, Lawrence Berkeley National Laboratory supported by the Office of Science, Office of Basic Energy Sciences, of the U.S. DOE under contract no. DE-AC02-05CH11231. Computer simulations used resources of the National Energy Research Scientific Computing Center (NERSC) through award BES-ERCAP-0028199, which is supported by the Office of Science of the U.S. Department of Energy, also under the same contract.

## Data and Code Availability

The entire SCOPE workflow can be downloaded from GitHub: https://github.com/atlas-nano/solvation_spectra.git.[39]

The sea_urchin code can be downloaded from GitHub: https://gitlab.com/electrolyte-machine/sea_urchin.git.[6]

## References

[1] J.N. Israelachvili, *Intermolecular and Surface Forces* (Academic press, 2011).
[2] D. Chandler, "Interfaces and the driving force of hydrophobic assembly," Nature **437**(7059), 640–647 (2005).
[3] J.D. Weeks, D. Chandler, and H.C. Andersen, "Role of Repulsive Forces in Determining the Equilibrium Structure of Simple Liquids," The Journal of Chemical Physics **54**(12), 5237–5247 (1971).
[4] K. Lum, D. Chandler, and J.D. Weeks, "Hydrophobicity at Small and Large Length Scales," J. Phys. Chem. B **103**(22), 4570–4577 (1999).
[5] K. Kobayashi, Y. Liang, T. Sakka, and T. Matsuoka, "Molecular dynamics study of salt–solution interface: Solubility and surface charge of salt in water," The Journal of Chemical Physics **140**(14), 144705 (2014).
[6] F. Roncoroni, A. Sanz-Matias, S. Sundararaman, and D. Prendergast, "Unsupervised learning of representative local atomic arrangements in molecular dynamics data," Phys. Chem. Chem. Phys. **25**(19), 13741–13754 (2023).
[7] F. Grasselli, "Investigating finite-size effects in molecular dynamics simulations of ion diffusion, heat transport, and thermal motion in superionic materials," The Journal of Chemical Physics **156**(13), 134705 (2022).
[8] E. Díaz-Herrera, E. Cerón-García, A. Bryan Gutiérrez, and G.A. Chapela, "Finite size effect on the existence of the liquid–vapour spinodal curve," Molecular Physics **120**(4), e1989071 (2022).

[9] J.M. Polson, E. Trizac, S. Pronk, and D. Frenkel, "Finite-size corrections to the free energies of crystalline solids," The Journal of Chemical Physics **112**(12), 5339–5342 (2000).

[10] B.M. Reible, J.F. Hille, C. Hartmann, and L. Delle Site, "Finite-size effects and thermodynamic accuracy in many-particle systems," Phys. Rev. Research **5**(2), 023156 (2023).

[11] G. Bussi, and A. Laio, "Using metadynamics to explore complex free-energy landscapes," Nat Rev Phys **2**(4), 200–212 (2020).

[12] A. Baskin, and D. Prendergast, "'Ion Solvation Spectra': Free Energy Analysis of Solvation Structures of Multivalent Cations in Aprotic Solvents," J. Phys. Chem. Lett. **10**(17), 4920–4928 (2019).

[13] A.P. Thompson, H.M. Aktulga, R. Berger, D.S. Bolintineanu, W.M. Brown, P.S. Crozier, P.J. In 'T Veld, A. Kohlmeyer, S.G. Moore, T.D. Nguyen, R. Shan, M.J. Stevens, J. Tranchida, C. Trott, and S.J. Plimpton, "LAMMPS - a flexible simulation tool for particle-based materials modeling at the atomic, meso, and continuum scales," Computer Physics Communications **271**, 108171 (2022).

[14] P. Mark, and L. Nilsson, "Structure and Dynamics of the TIP3P, SPC, and SPC/E Water Models at 298 K," J. Phys. Chem. A **105**(43), 9954–9960 (2001).

[15] I.S. Joung, and T.E. Cheatham, "Determination of Alkali and Halide Monovalent Ion Parameters for Use in Explicitly Solvated Biomolecular Simulations," J. Phys. Chem. B **112**(30), 9020–9041 (2008).

[16] R.W. Hockney, *Computer Simulation Using Particles* (crc Press, 2021).

[17] W. Shinoda, M. Shiga, and M. Mikami, "Rapid estimation of elastic constants by molecular dynamics simulation under constant stress," Phys. Rev. B **69**(13), 134103 (2004).

[18] G.J. Martyna, D.J. Tobias, and M.L. Klein, "Constant pressure molecular dynamics algorithms," The Journal of Chemical Physics **101**(5), 4177–4189 (1994).

[19] M. Parrinello, and A. Rahman, "Polymorphic transitions in single crystals: A new molecular dynamics method," Journal of Applied Physics **52**(12), 7182–7190 (1981).

[20] M.E. Tuckerman, J. Alejandre, R. López-Rendón, A.L. Jochim, and G.J. Martyna, "A Liouville-operator derived measure-preserving integrator for molecular dynamics simulations in the isothermal–isobaric ensemble," J. Phys. A: Math. Gen. **39**(19), 5629–5651 (2006).

[21] G. Fiorin, M.L. Klein, and J. Hénin, "Using collective variables to drive molecular dynamics simulations," Molecular Physics **111**(22–23), 3345–3362 (2013).

[22] G. Fiorin, J. Hénin, and A. Kohlmeyer, "Collective variables module reference manual for LAMMPS," (2013).

[23] P. Raiteri, A. Laio, F.L. Gervasio, C. Micheletti, and M. Parrinello, "Efficient Reconstruction of Complex Free Energy Landscapes by Multiple Walkers Metadynamics," J. Phys. Chem. B **110**(8), 3533–3539 (2006).

[24] M. Invernizzi, and M. Parrinello, "Rethinking Metadynamics: From Bias Potentials to Probability Distributions," J. Phys. Chem. Lett. **11**(7), 2731–2736 (2020).

[25] T.M. Schäfer, and G. Settanni, "Data Reweighting in Metadynamics Simulations," J. Chem. Theory Comput. **16**(4), 2042–2052 (2020).

[26] A.J. Fontana, "Appendix A: Water Activity of Saturated Salt Solutions," in *Water Activity in Foods*, edited by G.V. Barbosa-Cánovas, A.J. Fontana, S.J. Schmidt, and T.P. Labuza, 1st ed., (Wiley, 2007), pp. 391–393.

[27] C. Monnin, M. Dubois, N. Papaiconomou, and J.-P. Simonin, "Thermodynamics of the LiCl + $H_2O$ System," J. Chem. Eng. Data **47**(6), 1331–1336 (2002).

[28] A. France-Lanord, and J.C. Grossman, "Correlations from Ion Pairing and the Nernst-Einstein Equation," Phys. Rev. Lett. **122**(13), 136001 (2019).

[29] Ø. Gullbrekken, I.T. Røe, S.M. Selbach, and S.K. Schnell, "Charge Transport in Water–NaCl Electrolytes with Molecular Dynamics Simulations," J. Phys. Chem. B **127**(12), 2729–2738 (2023).

[30] K.D. Fong, J. Self, B.D. McCloskey, and K.A. Persson, "Ion Correlations and Their Impact on Transport in Polymer-Based Electrolytes," Macromolecules **54**(6), 2575–2591 (2021).

[31] T.A. Pascal, and W.A. Goddard, "Interfacial Thermodynamics of Water and Six Other Liquid Solvents," J. Phys. Chem. B **118**(22), 5943–5956 (2014).

[32] P.G. Bolhuis, C. Dellago, P.L. Geissler, and D. Chandler, "Transition path sampling: throwing ropes over mountains in the dark," J. Phys.: Condens. Matter **12**(8A), A147–A152 (2000).

[33] C. Dellago, P.G. Bolhuis, and D. Chandler, "Efficient transition path sampling: Application to Lennard-Jones cluster rearrangements," The Journal of Chemical Physics **108**(22), 9236–9245 (1998).

[34] Y. Luo, and B. Roux, "Simulation of Osmotic Pressure in Concentrated Aqueous Salt Solutions," J. Phys. Chem. Lett. **1**(1), 183–189 (2010).

[35] S.-T. Lin, P.K. Maiti, and W.A. Goddard, "Two-Phase Thermodynamic Model for Efficient and Accurate Absolute Entropy of Water from Molecular Dynamics Simulations," J. Phys. Chem. B **114**(24), 8191–8198 (2010).

[36] B. Hess, C. Holm, and N. Van Der Vegt, "Osmotic coefficients of atomistic NaCl (aq) force fields," The Journal of Chemical Physics **124**(16), 164509 (2006).

[37] X. Cao, H. Jia, W. Xu, and J.-G. Zhang, "Review—Localized High-Concentration Electrolytes for Lithium Batteries," J. Electrochem. Soc. **168**(1), 010522 (2021).

[38] V. Alexandrov, U. Lehnert, N. Echols, D. Milburn, D. Engelman, and M. Gerstein, "Normal modes for predicting protein motions: A comprehensive database assessment and associated Web tool," Protein Science **14**(3), 633–643 (2005).

[37] X. Ruan, "solvation_spectra," atlas-nano/solvation_spectra: v1.0.1 (v1.0.1). Zenodo (2025). https://doi.org/10.5281/zenodo.15377288.